# SECURING CLOUDS – THE QUANTUM WAY

*Marmik Pandya*
*Department of Information Assurance*
*Northeastern University*
*Boston, USA*

## 1 Introduction

Quantum mechanics is the study of the small particles that make up the universe – for instance, atoms et al. At such a microscopic level, the laws of classical mechanics fail to explain most of the observed phenomenon. At such a state quantum properties exhibited by particles is quite noticeable.

Before we move forward a basic question arises that, what are quantum properties? To answer this question, we'll look at the basis of quantum mechanics – The Heisenberg's Uncertainty Principle. The uncertainty principle states that the more precisely the position is determined, the less precisely the momentum is known in this instant, and vice versa. For instance, if you measure the position of an electron revolving around the nucleus an atom, you cannot accurately measure its velocity. If you measure the electron's velocity, you cannot accurately determine its position.

Equation for Heisenberg's uncertainty principle:

$$\sigma_x \sigma_p \geq \frac{\hbar}{2}$$

Where $\sigma_x$ standard deviation of position, $\sigma_x$ the standard deviation of momentum and $\hbar$ is the reduced Planck constant, $h / 2\pi$.

In a practical scenario, this principle is applied to photons. Photons – the smallest measure of light, can exist in all of their possible states at once and also they don't have any mass. This means that whatever direction a photon can spin in -- say, diagonally, vertically and horizontally -- it does all at once. This is exactly the same as if you constantly moved east, west, north, south, and up-and-down at the same time.

Quantum entanglement is a physical phenomenon that occurs when pairs or groups of particles are generated or interact in ways such that the quantum state of each particle cannot be described independently—instead, a quantum state may be given for the system as a whole. Consider the scenario in which a physical process creates a pair of photon such that the total spin of the system is null. Now, if a photon is examined by a human observer after it has already traveled a million light year, and its spin is vertical. Then it is certain that another photon which is two million light years away from the first one at that point will have horizontal spin. Up to the point of measurement, the polarization of both photons is unknown. It is hard to believe, but the act of measurement will actually cause the other photon to commit to a certain state. Many experiments have proved this concept. Einstein's famous quote, "God does not play dice with the universe" was a comment on the bizarre effects of quantum mechanics.

Nonetheless, we do not know everything there is to know about this complex science. In the words of physicist Richard Feynman: "I can safely say that nobody understands quantum mechanics". This uncertainty partially explains the difficulty engineers and scientists alike have encountered in the task of building a quantum computer.

## 2 Quantum Cryptography

CIA triad – Confidentiality, Integrity, and Availability are basic goals of security architecture. To ensure CIA, many authentication scheme has been introduced in several years. Currently deployment of Public Key Infrastructure (PKI) is a most significant solution. PKI involving exchange key using certificates via a public channel to a authenticate users in the cloud infrastructure. However, there is a certain issue pertaining to the PKI authentication where the public key cryptography only provide computational security because PKI is based on Asymmetric Key Cryptography. It is exposed to widespread security threats such as eavesdropping, man in the middle attack, masquerade et al. This paper aims to look into basic security architecture in place currently and further it tries to introduce a new proposed security architecture, which makes use of the knowledge of Quantum Mechanics and current advances in research in Quantum Computing, to provide a more secure architecture.

**2.1 Importance of Qubits**

In a classical computing system, a bit would have to be in one state or the other. However in a quantum computing system, quantum mechanics allows the qubit to be in a superposition of both states at the same time. In quantum computing, a **qubit** or **quantum bit** (sometimes **qbit**) is a unit of quantum information—the quantum analogue of the classical bit. A qubit is a two-state quantum-mechanical system, such as the polarization of a single photon: here the two states are vertical polarization and horizontal polarization.

| System | Qubit State |
|---|---|
| Electron | Spin |
| Photon | Polarization |

Table 1: Examples of Physical Qubits

A qubit has a few similarities to a classical bit but is overall very different. There are two possible outcomes for the measurement of a qubit—usually 0 and 1, like a bit. The difference is that whereas the state of a bit is either 0 or 1, the state of a qubit can also be a superposition of both.[4] It is possible to fully encode one bit in one qubit. Hence, a qubit can hold even more information, e.g. up to two bits using superdense coding.

**2.2 Shor's Algorithm**

In 1994, Shor proposed an algorithm for period finding and then subsequently integer factorization problem. Later, Shor also proposed an efficient quantum algorithm for the discrete logarithm problem. Shor's algorithm consists of Classical Part and Quantum Part. Quantum part of the algorithm, uses quantum Fourier transform to find the period of a certain function, which is infeasible with classical computers, but in 2001 a group at IBM, who factored 15 into 3 × 5, using an NMR implementation of a quantum computer with 7 qubits.

Shor mathematically showed that the quantum part runs in time **O ((log n) $^2$ (log log n)(log log log n))** on a quantum computer. Next, it must perform **O (log n)** steps of post processing on a classical computer to execute the continued fraction algorithm.

Factorizations and discrete logarithm problem are two of the most difficult problems arising in the breaking of current cryptographic algorithms. If the Shor's algorithm is implemented on Quantum Computers, no application using this algorithm will be able to withstand the attackers.

| System | Underlying Hard Problem |
|---|---|
| RSA | Factorization |
| Rabin's Cryptosystem | Factorization |

| KMOV | Factorization |
|---|---|
| Diffie-Hellman Key Exchange | Discrete Logarithm Problem |
| El Gamal | Discrete Logarithm Problem |
| Elliptic Curve Cryptography (ECC) | Discrete Logarithm Problem |
| Digital Signature Algorithm (DSA) | Discrete Logarithm Problem |

**Table 2: Cryptosystems broken by Shor's algorithm**

**2.3 Quantum Cryptography**

The uncertainty principle, possible of indivisible quanta and the quantum entanglement forms the basis of the quantum cryptography. The no-cloning theorem, presented by Wootters and Zurek in 1982, forms another basis of Quantum Cryptography. As a direct application of no cloning theorem – Eavesdropper cannot interpret the unknown qubits i.e. the unknown quantum states, which makes the use of qubits in key transmission for asymmetric cryptography resistant to man in the middle attack. Hence, it is attracting considerable attention as a replacement for other contemporary cryptographic methods, which are based on computational security.

Quantum Cryptography doesn't reinvent the wheel as a whole. Internally, it works just like a traditional asymmetric cryptographic system. But while cryptographic methods like RSA, use computational difficulty in breaking the key, Quantum Cryptographic system that uses quantum physics for key transmission. Quantum cryptographic transmission encrypts the 0s and 1s of a digital signal on individual particles of light - photons. Each type of a photon's spin represents one piece of information - usually a 1 or a 0, for binary code. This code uses strings of 1s and 0s to create a coherent message. For example, 01101000 01101001 could correspond with "hi". Now, a binary code can be assigned to each photon -- for example, a photon that has a **vertical spin ( | )** can be assigned a 1 and a photon with a horizontal spin, can be assigned 0. Alice can send her photons through randomly chosen filters and record the polarization of each photon. She will then know what photon polarizations Bob should receive. Now, even if eve detects (eavesdrops on) the signal, the information on the photons is suddenly transformed, meaning both that it is immediately noticeable that eavesdropping has appeared and that the third party is not able to decrypt the information.

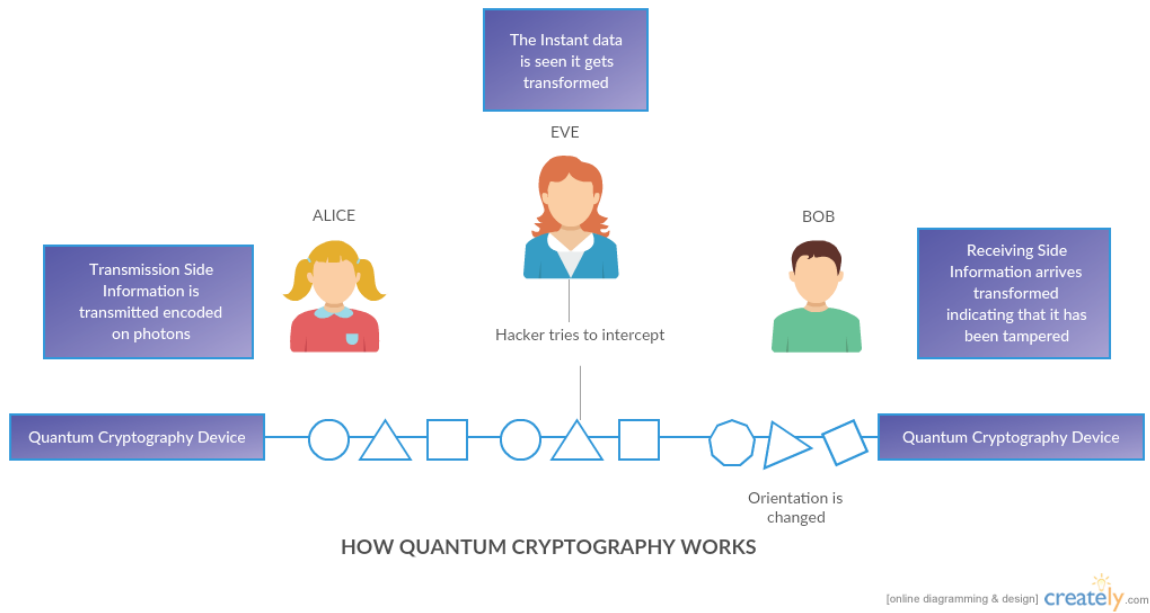

**Figure 1: The working of Quantum Cryptography**

### 2.3.1 Quantum Key Distribution (QKD)

Quantum Key Distribution is the most famous application of Quantum Cryptography. Before understanding the basics of QKD, let's see current encryption standards. Currently, in public key cryptography, before transferring data, both Alice and Bob agrees upon a shared secret key. Alice uses the public key of Bob to transfer the shared secret key to Bob and that encrypted key can be decrypted only by Bob's private key. Now, Bob uses his private key to decrypt the shared key and then using that shared secret key, Bob can decrypt all the encrypted messages that Alice sends. This type of system is susceptible to Man in the Middle attack since the assumption used for transmission of shared key is that decrypting it without the key is, computationally infeasible. But with Shor's algorithm, even this isn't computationally infeasible anymore.

This is where QKD walks in. The Quantum Key Distribution is a method used in the framework of quantum cryptography in order to produce a perfectly random key which is shared by a sender and a receiver while making sure that nobody else has a chance to learn about the key, e.g. by capturing the communication channel used during the process. The best known and popular scheme of quantum key distribution is based on the Bennet–Brassard protocol (i.e. BB84). It depends on the no-cloning theorem for non-orthogonal quantum states.

The basic principle of the Quantum Key Distribution (QKD) using the BB84 protocol, involves sending decryption keys as quantum particles. Thanks to the quantum properties of these particles, sender, and the receiver can surely identify if their communication was subjected to man in the middle attack.

To detect the intruders, the photons can be randomly sampled for different properties. Now, since the measurement in one property results in uncertainty in the measurement of other property, Alice and Bob independently chooses to measure each proton for different properties, say polarization or spin. They then exchange which property they measured on each photon, and examine whether the values are the same on photons that they measured are same or not. If there is a large difference, it is likely the signal was intercepted, and the communication should be dropped. If results are similar, then the values can be stored as binary data; for instance, left spin = 0, right spin = 1. This is the shared key.

Once both Alice and Bob have agreed upon the shared secret key, they use the normal channel to transfer the data encrypted with the shared key.

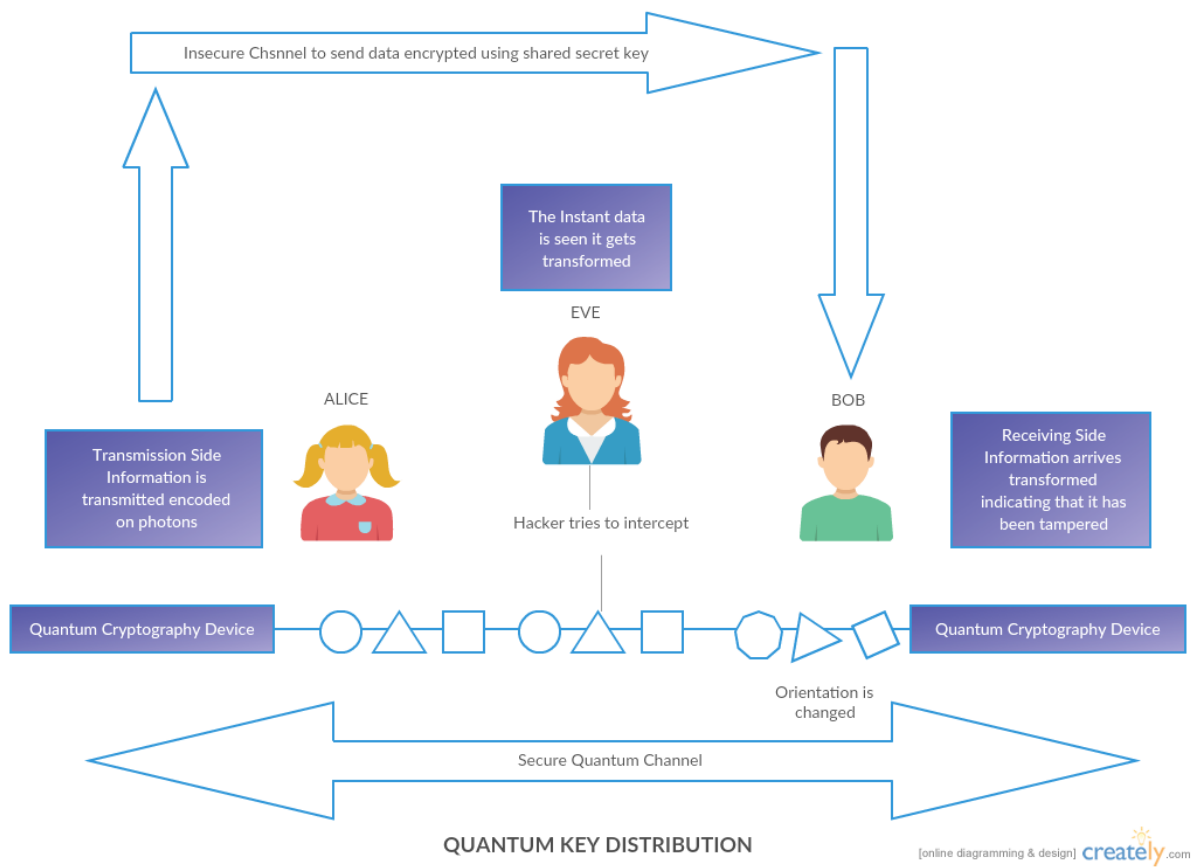

**Figure 2 – Quantum Key Distribution**

# 3 Application of QKD into Public Key Cryptography

**3.1 PUBLIC KEY CRYPTOGRAPHY USING RSA ALGORITHM**

Factoring is the underlying presumably hard problem upon which several public-key cryptosystems are based.

Factoring is widely believed to be a hard problem and the best algorithm for solving it is the Number Field Sieve with a sub-exponential running time. The principal threat comes from a quantum computer on which factoring can be solved efficiently using Shor's algorithm. The most popular cryptosystem based on factorization is RSA. RSA was invented by Rivest, Shamir and Adelman in 1978. It can be summarized as follows:

**1. Key generation:**

- Choose two large primes p and q and compute the RSA modulus N = pq.
- Choose an integer e that is coprime to (p – 1) (q – 1).
- Compute d using ed ≡ 1 (mod (p – 1) (q – 1)).
- Publish the public key (N, e) and keep the private key (N, d).

**2. Encryption:**
- Represent the message to be transmitted as a positive integer m < N.
- Encrypt m with the public key (N, e) using c ≡ me (mod N).

**3. Decryption:**
- The receiver decrypts the message using m ≡ c d (mod N).
- Transform the positive integer m into the original message.

The idea of breaking RSA with a quantum computer using Shor's algorithm was a powerful motivator for the design and construction of quantum computers and for the study of new quantum computer algorithms and cryptosystems that are secure from quantum computers.

**3.2 IMPLEMENTING RSA ALGORITHM WITH QKD**

Factorization is the underlying hard part in RSA algorithm. Using the Shor's algorithm, it is easy to break the message that has been encrypted using RSA. So, here we can introduce QKD for key generation and distribution and use underlying principles of RSA to transmit the message.

Following are the steps for the new improvised RSA.

1. Alice sends a request to QKD to initiate a conversation with Bob.
   Alice → QKD: $E_{PR\text{-}ALICE}$ ( $ID_{ALICE}$ || $ID_{BOB}$ ).

2. QKD logs the Alice's Request and notifies Bob about the possible connection
   QKD → BOB: $E_{PU\text{-}BOB}$ ( $ID_{ALICE}$ || $ID_{BOB}$ )

3. Bob replies by accepting the connection.
   Bob → QKD: $E_{PR\text{-}BOB}$ ( $ID_{ALICE}$ || $ID_{BOB}$ )

4. QKD creates a session key using quantum bases (+, X) in some order and starts distributing those to Alice and Bob, Alice and Bob will use those bases to communicate.
   QKD → Alice: $E_{PU\text{-}ALICE}$ ( $ID_{ALICE}$ || $ID_{BOB}$ || SK )
   QKD → Alice: $E_{PU\text{-}BOB}$ ( $ID_{ALICE}$ || $ID_{BOB}$ || SK )

5. Alice encrypts the message using the session key and sends it to Bob over a Quantum Channel. Next, Alice also sends random bits to QKD.
   Alice → BOB: $E_{PR\text{-}ALICE}$ ( $E_{SK}$ (Message) || $ID_{BOB}$ )

6. Bob decrypts the message using the Session Key and sends the random bits to QKD.

7. QKD checks the random bits to know if there's any intruder. If there's an intruder QKD notifies Alice and Bob and discards the Session Key to create a new Session Key.

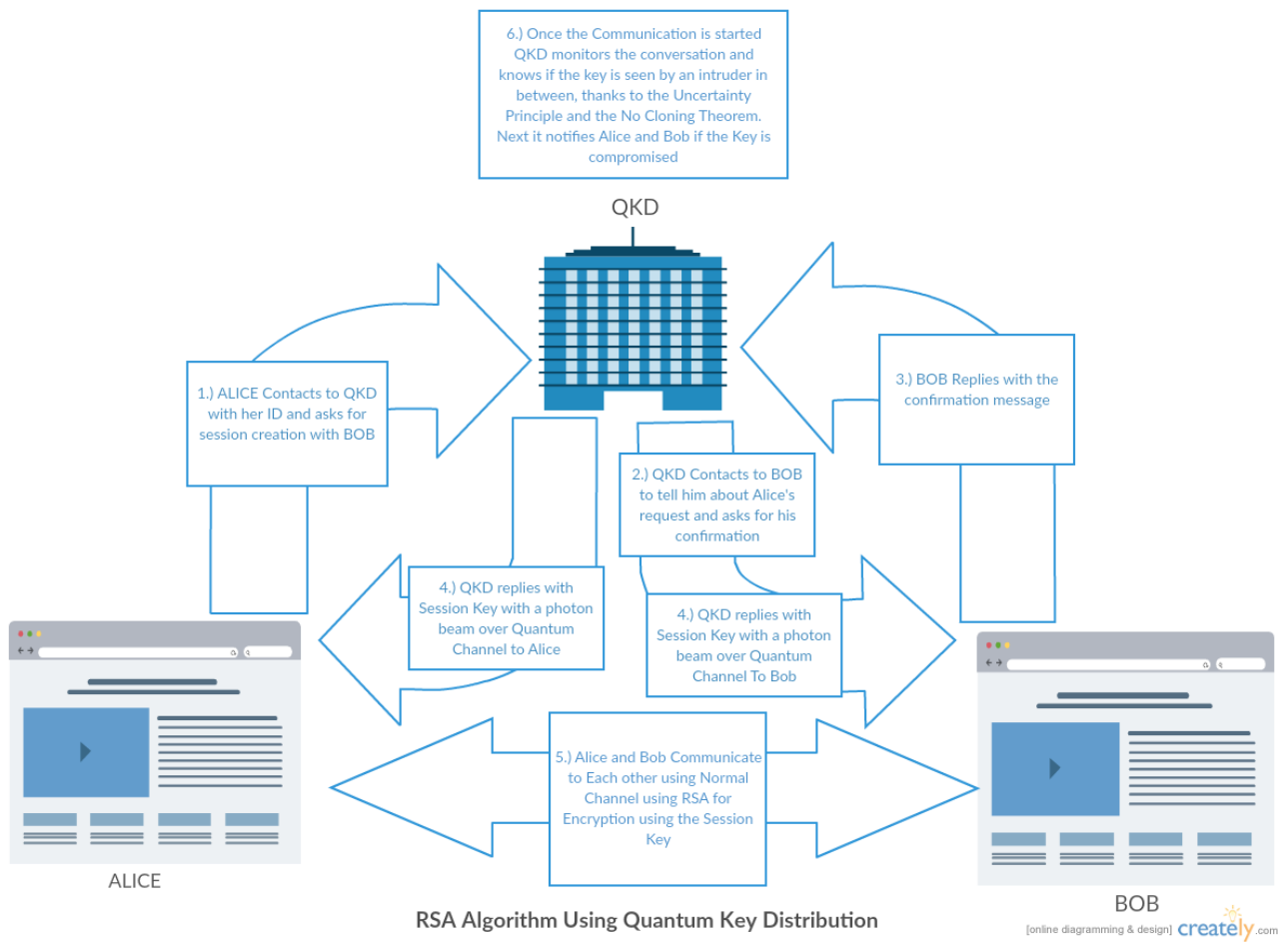

**Figure 3: RSA Algorithm Using Quantum Key Distribution**

# 4 Secure Cloud Computing

With the success of Internet, there has been a rapid and significant success in the development of data processing and data storage technologies. These advancements in storage techniques alongside SaaS techniques have enabled a different computing model – Cloud Computing. Examples of such service providers include big players like Google, Microsoft, Apple, Amazon et al. Since the data transfer for such an application occurs through the classical network, storage on the same server for many users where resource allocation and scheduling is provided by the cloud service provider and with the breakthrough in malicious programs, cloud security becomes an important issue. Every day hackers are trying to hack into some cloud or the other and recently with the security of giants like Apple and Dropbox being compromised, cloud security has become a hot topic. Here, I'm trying to propose a new hybrid security architecture for the cloud which uses benefits of current protocols like Kerberos and security benefits of Quantum Cryptography.

**4.1 Current Work**

Currently, deploying public key infrastructure (PKI) is one of the most elegant solutions for securing clouds. PKI works by exchanging keys using certificate via the public channel for authentication purposes. PKI based architecture works on public key cryptography. As seen before public key cryptography provides only with the computational security and with the arrival of Quantum Computers, using Shor's algorithm, it'll be easy to break the public key cryptographic algorithms. Apart from this, Asymmetric Key Cryptography is also vulnerable to many kinds of security threats like the man in the middle attack, masquerade et al. Apart from this PKI infrastructure is too complex, involving a lot of entities which is very difficult to deploy.

Another method for secure authentication on Cloud is using Kerberos. Many researchers have proposed a Kerberos-based model for secure data storage and secure authentication on the cloud. There are many benefits for using Kerberos in cloud computing, with the major one being the property of Kerberos that allows the nodes to connection points of the various cloud networks, and to communicate with each other. Apart from this compared to PKI, Kerberos is easy to deploy and it uses a session key which enables the possibility of Single Sign On.

Summary of Kerberos Message Exchange in Cloud Service:

| |
|---|
| **(A) AS Exchange: to obtain TGT** |
| 1. AS_REQ – {cloud customer name, expiration time, tgs cloud service name, …} |
| 2. AS_REP – {$S_{A, KDC}$, expiration time, tgs cloud service name …}. $K_A$ + {$S_{A, KDC}$, expiration time, cloud customer name …}. $K_{KDC}$. |
| **(B) Ticket Granting Sever Exchange: to obtain Server Granting Ticket** |
| 3. TGS_REQ – {timestamp, checksum …}.$S_{A, KDC}$ + { $S_{A,KDC}$, expiration time , cloud customer name, …}. $K_{KDC}$ + cloud service name + expiration time |
| 4. TGS_REP – {$S_{A,B}$ , cloud service name, expiration time, …}.$S_{A, KDC}$ + {$S_{A, B}$ ,cloud customer name, expiration time,…}. $K_B$ |
| **(C) Customer/Server Authentication Exchange: to obtain Cloud Service** |
| 5. CS_REQ – {timestamp, checksum …}.$S_{A,B}$ + {$S_{A,B}$ , cloud customer name, expiration time, …}. $K_B$ |
| 6. CS_REP – {timestamp}.$S_{A,B}$ |

**Figure 4: Cloud Computing with Kerberos authentication [Source: Yaser Fuad Al-Dubai and 2Dr. Khamitkar S.D , 2013, A PROPOSED MODEL FOR DATA STORAGE SECURITY IN CLOUD COMPUTING USING KERBEROS AUTHENTICATION SERVICE]**

**4.2 Proposed Model**

As shown above, implementation of Kerberos model to cloud computing, is a very advantageous. But Kerberos also uses algorithms like DES and AES to generate the key. With the help of Groover's Algorithm, searching an unsorted database with N entries in $O(\sqrt{N})$ time rather than the usual $O(N)$ time. For AES-256, it currently takes an average of n/2 guesses to break, i.e. $2^{255}$. However with quantum computers this can be done in $2^{128}$ time, which is very much faster. Now that's only the brute force for AES-256, with the cleverer attacks like using rainbow tables, it can be broken even faster. Hence, for the Post Quantum world, using QKD for Key Generation and Key Distribution within KDC alongside classical Kerberos implementation, would result in better security. Also with the world moving towards the Internet of Things (IoT) Revolution, Single Sign On solution provided by the Kerberos Model could be the ideal security solution. Also, at the brink of this IoT revolution, security issues are of great concern and the possibility of Quantum Computers in near future would mean, when they arrive nothing will be secure. Hence, using QKD inside of the KDC could be the ideal solution.

For the basic approach for cloud computing with Kerberos authentication, the given architecture is almost similar to the Architecture mentioned in the previous section. The only difference is that the Data Transmission between the Cloud and the Client happens through Quantum Channel and there is a QKD inside the KDC.

Now, to use the services of the cloud, a cloud customer should supply a ticket. A ticket for a cloud service is a series of bits with the attribute that it has been enciphered using the private key for that cloud service. That Session key is stored in the global database shared between the cloud service itself and the Kerberos. Only KDC has the write access to the Database so the cloud service can be confident that any information that exists within the database regarding ticket originated from Kerberos. Kerberos will have placed the identity of the cloud customer inside the database matched with the Session Key, so the cloud service that receives a ticket can look up to the database to find the session key to decrypt the data. To help ensure that one customer does not steal and reuse another customer's tickets, the cloud customer accompanies the ticket with an authenticator. (In addition, tickets expire after a specified lifetime, which is usually within a few hours.). The cloud customer gets a ticket by sending a message to KDC naming the principal identifier of the desired cloud service, the principal identifier of the (alleged) cloud customer and the reference to the current time of day. In response, the KDC authenticates the sender since the request is encrypted using his password. Once the client is authenticated, KDC replies back with the Ticket granting Ticket. Anyone can intercept the message and get the ticket granting ticket, but the TGT has the client identification embedded inside and it is encrypted using the KDC's private key. The cloud customer will be able to unseal this message, obtain the ticket. No other customer without the named cloud customer's private key [password] can correctly decrypt the reply to produce the sealed tickets and corresponding session key.

Once a cloud customer gets a ticket and wants to use the cloud service, it generates a random quantum bases and sends it to a KDC along with the ticket. Now, anyone can intercept this message but it is of no use to them since they don't have the password to decrypt and even after the brute force attack, to generate the session key they require server's Quantum Bases. In return, KDC generates the Session key and stores it in the database and sends back the Server's Quantum Bases to the Client via a quantum channel. If someone intercepts, the bases change and the session key that client will generate will not match to the session key in the database, which will not let the communication go through.

The client computes the session key using it's Quantum Bases and Server's Quantum Bases and then uses the quantum channel to transmit data. Cloud service can identify the client in the database using client ID and then get the session key to decrypt the message. If the session key cannot decrypt the message Cloud service provider can conclude that somewhere along the line, there is an intruder and entire will have to be repeated again

**Steps:**

- First the Cloud Service Provider generates random Quantum base and shares it with KDC.

- When a Client logs in, it first sends the request containing Client Name et al. to the KDC encrypted with its own password using the classical channel.
  Client → KDC: $E_{PASSWORD\text{-}CLIENT}$ (Client Address) || $ID_{CLIENT}$.

- Authentication Server inside the KDC authenticates the client and sends it the ticket-granting ticket (TGT).
  KDC → Client: $E_{PASSWORD\text{-}CLIENT}$ (TGT).

- When a client wants to access the cloud, it generates the random quantum base and sends it to the KDC along with TGT encrypted with its own password via the classical channel.
  Client → KDC: $E_{PASSWORD\text{-}CLIENT}$ ($QB_{CLIENT}$ || TGT) || $ID_{CLIENT}$.

- KDC generates a session key by comparing the quantum bases of the cloud service provider and client and stores the session key in the global database.

- After KDC generates the session key, it communicates the base to the client via quantum channel, due to which client can compute the session key itself using the it's base.
  KDC → Client: $E_{PASSWORD\text{-}CLIENT}$ ($QB_{CLOUD\text{-}SERVICE\text{-}PROVIDER}$).

- Once, the client computes the session key, it uses that session key to encrypt and send data to the server via a quantum channel. The client doesn't encrypt its Client ID since server uses the client ID to find the session key to decrypt the data.
  Client → Service Provider: $E_{SESSION\text{-}KEY}$ (FILE) || $ID_{CLIENT}$.

The major advantage of using this model is that it uses ease of deployment of a Kerberos Model along with the security benefits of Quantum Cryptography. Kerberos allows the nodes to connection points of the various cloud networks to communicate with each other which in turns helps in providing Single Sign-On solution for using various cloud networks by signing on only once. Any Encryption Algorithm AES, DES et al. can be used for encrypting data.

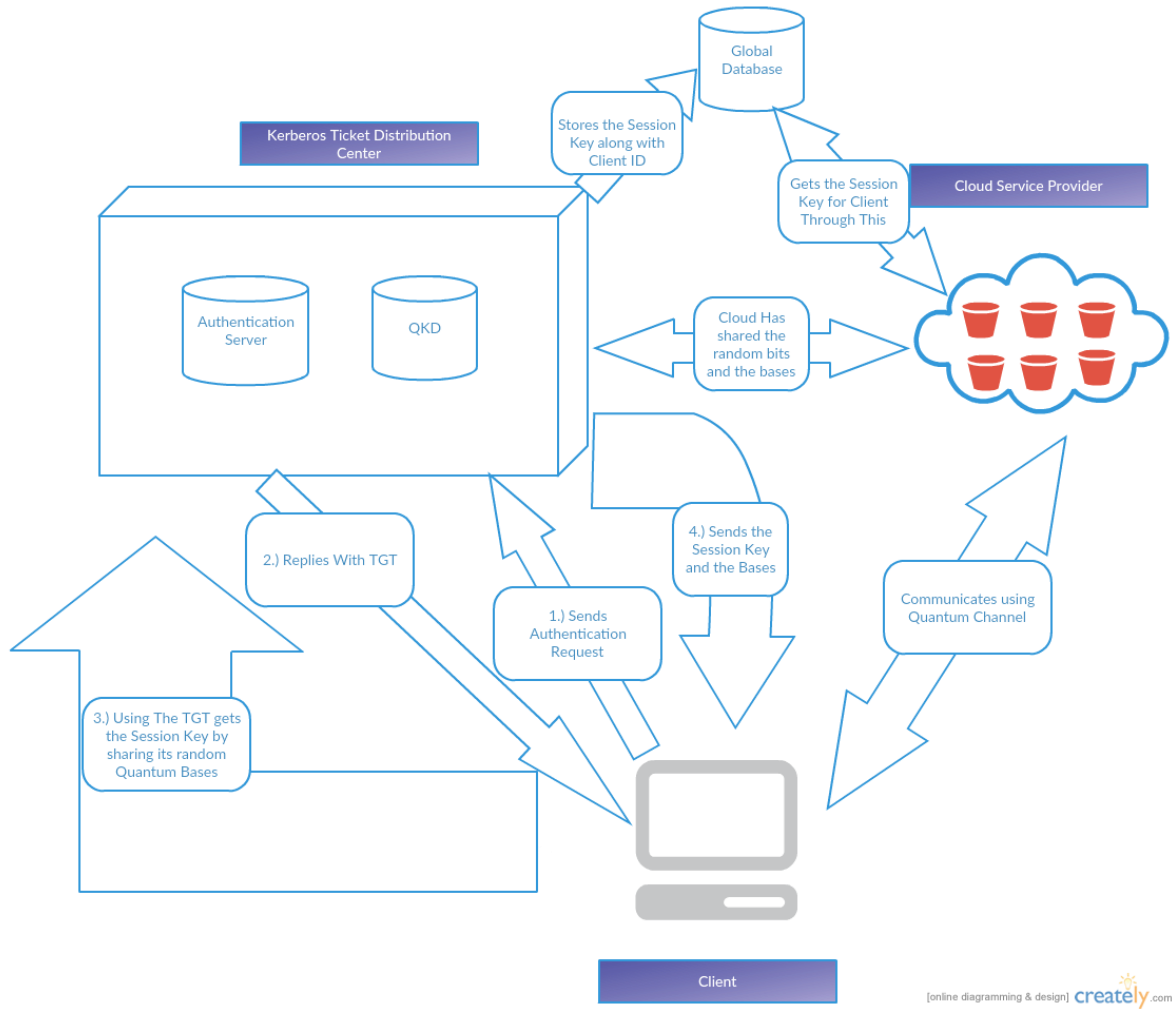

**FIGURE 5: Cloud Computing with Kerberos and QKD based authentication and Key Distribution**

# 5  Conclusion and Future Works

In a nutshell, this paper has introduced a new security architecture for Cloud Computing. This new method builds on top of the pre-existing architecture of using Kerberos for Single Sign-On authentication for flexibility and scalability but gives a workaround for the limitation of classical cryptographic algorithms by using QKD inside the KDC for key distribution and using Quantum Channel for transmission. This paper introduced a new cloud computing environment, which suggested integrates and uses ease and simplicity of Classical Cryptography models and secure benefits of QKD as a new hybrid technique. Compared to current models, my attempt is better than existing models in following ways:

1. Gives the flexibility and the scalability of an ideal Kerberos-based solution
2. QKD based method for sharing keys is more secure than existing cloud computing architecture deployed upon AES and/or PKI.
3. Since, there's less computation included compared to PKI or AES for key generation, it's faster than existing models.

Since, "There's always a hack for everything", in future potential attacks on this like Large Pulse attack, Time Shift Attack, Fake State Attacks et al. can be studied into and workarounds can be addressed. Apart from this in future, although tremendously complex, proper deployment architecture and the statistical evidence of the benefits could be studied.